\documentclass[aps,twocolumn,showpacs,pra,amsmath,amssymb,floatfix,superscriptaddress]{revtex4-1}
\usepackage{amsmath,amssymb,graphicx,color}
\usepackage{bm}
\usepackage{bbm}
\usepackage[caption=false]{subfig}  
\usepackage[usenames,dvipsnames]{xcolor}
\usepackage[normalem]{ulem}
\usepackage{soul}
\usepackage[colorlinks=true,citecolor=Cerulean,linkcolor=RubineRed,urlcolor=Cerulean]{hyperref}
\usepackage{soul,xcolor}

\newcommand{\Tr}{\ensuremath{\mathrm{Tr}\,}}
\newcommand{\avg}[1]{\ensuremath{\langle #1 \rangle}}

\usepackage{mathptmx} 
\usepackage{braket}

\begin{document}
\title{Quantum dynamics under simultaneous and continuous measurement of noncommutative observables}

\author{Chao Jiang} \email{chaojiang@zju.edu.cn}
\affiliation{Department of Physics and Zhejiang Institute of Modern Physics, Zhejiang University, Hangzhou, Zhejiang 310027, China}

\author{Gentaro Watanabe}\email{gentaro@zju.edu.cn}
\affiliation{Department of Physics and Zhejiang Institute of Modern Physics, Zhejiang University, Hangzhou, Zhejiang 310027, China}
\affiliation{Zhejiang Province Key Laboratory of Quantum Technology and Device, Zhejiang University, Hangzhou, Zhejiang 310027, China}

\begin{abstract}
We consider simultaneous and continuous measurement of two noncommutative observables of the system whose commutator is not necessarily a $c$-number. We revisit the Arthurs-Kelly model and generalize it to describe the simultaneous measurement of two observables of the system. Using this generalized model, we continuously measure the system by following the scheme proposed by Scott and Milburn [Scott and Milburn, \href{https://doi.org/10.1103/PhysRevA.63.042101}{Phys. Rev. A {\bf 63}, 042101 (2001)}]. We find that the unconditioned master equation reduces to the Lindblad form in the continuous limit. In addition, we find that the master equation does not contain a cross term of these two measurements. Finally, we propose a scheme to prepare the state of a two-level system in an external field by feedback control based on the simultaneous, continuous measurement of the two observables.
\end{abstract}

\maketitle

\section{Introduction \label{sec:intro}}  

Quantum feedback control \cite{H. M. W., K. Jacobs2, H. M. Wiseman,  Diosi94, S. Lloyd, Daniel, Sagawa, Brif} is a widely employed technique to drive a quantum system to a desired state \cite{Bushev, Dotsenko, Gillett, Vijay}.  Using measurement results to control system parameters, the feedback control technique provides a robust way to prepare the target state without fine tuning the protocol. The first application of quantum feedback control dates back to a scheme proposed by Yamamoto and his collaborators to produce an amplitude-squeezed state \cite{Yamamoto1, Yamamoto2}. In the recent decades, the development of cavity quantum electrodynamics (cavity QED) \cite{J. M. Raimond, H. Mabuchi, A. Blais} and circuit quantum electrodynamics (cQED) based on superconducting qubits \cite{A. Wallraff, J. Clarke} has offered a promising platform for quantum information and computation. As a consequence, feedback control technology finds a broad application in the new field such as qubit resetting \cite{D. Riste1}, state stabilization \cite{C. Sayrin}, quantum error correction \cite{J. Cramer}, entanglement enhancement \cite{D. Riste2}, etc.

In the standard protocol of feedback control of a quantum system, continuous measurement \cite{V. B. Braginsky, T. A. Brun, K. Jacobs, H. M. W., K. Jacobs2, A. A. Clerk} is performed and the state of the system is manipulated in parallel according to the measurement outcome. Continuous measurement can be realized by coupling the system to an auxiliary system working as a measurement apparatus. By continuously interacting with the system, the state of the apparatus is influenced by the system, and hence the information of the system can be extracted continuously \cite{H.-P. Breuer, H. M. W., K. Jacobs2}. If the coupling is weak enough and the apparatus does not have memory, i.e., the so-called Born-Markov approximation, the master equation of the system can be reduced to the celebrated Lindblad-Gorini-Kossakowski-Sudarshan form (in short, Lindblad form hereafter) \cite{H.-P. Breuer, H. M. W., K. Jacobs2, Lindblad, Gorini76, Carmichael}.

Continuous measurement of a single observable has already been discussed in many works \cite{Barchielle, N. Gisin, C. M. Caves, L. Diosi, Presilla, A. Steck, J. Combes, T. Konrad, C. Laflamme, Yang}, and the feedback control theory based on the single-observable measurement is well established \cite{A. C. Doherty1, A. C. Doherty2}. On the other hand, as for simultaneous and continuous measurement of multiple observables, there is another fundamental issue when two noncommutative observables are simultaneously measured. Although several works have discussed the simultaneous, continuous measurement of two noncommutative observables, they either consider two particular observables, such as canonically conjugate variables \cite{A. J. Scott, Gough, Ochoa} (i.e., the commutator of them is a $c$-number) and qubit observables \cite{Chantasri, Luis, Shay}, or assume that the system under the simultaneous and continuous measurement evolves according to the Lindblad form master equation without additional cross terms, which describe the interplay effect of the two individual measurements, even though the measured observables are not commutative \cite{K. Jacobs2, Chantasri, A. Levy}. In the present paper, instead of assuming, we will derive the master equation of the system under the simultaneous, continuous measurement of two arbitrary noncommutative observables, whose commutator is not a $c$-number but an operator. We will start from a concrete measurement model, which can be used to describe the simultaneous and continuous measurement of two arbitrary observables, and show that the master equation obtained for simultaneous and continuous measurement of two observables whose commutator is a $c$-number \cite{A. J. Scott} is still valid regardless of the observables to be measured.

State preparation of qubits \cite{L. DiCarlo, C. Song, N. Friis, T. L. Patti} has always been a crucial issue in quantum information processing and quantum computation. This issue becomes more practically important in the current situation in which the recent development of cavity QED and cQED has spurred the fabrication of quantum computers. In addition, regarding our current problem, the angular momentum operators in a two-level system (TLS) of the qubit [i.e., spin-$1/2$ operators] are one of the simplest but nontrivial examples of observables whose commutator is not a $c$-number. Among various techniques, the feedback control is a promising scheme that allows us to control the qubit state in a robust manner. Further, feedback control using the measurement outcome of two observables instead of one provides us with a more flexible way to control the system. Therefore, preparation of a designated target state of a TLS based on the simultaneous, continuous measurement of two noncommutative observables is a challenging but important task.

In this paper, we derive the unconditioned and conditioned master equations of the system under the continuous and simultaneous measurement, and we also provide a state preparation scheme for a TLS using a static external field, simultaneous and continuous measurement, and feedback control as an application of our formalism. First, we generalize the Arthurs-Kelly measurement model \cite{E. Arthurs, S. L. Braunstein, A. J. Scott} to the simultaneous measurement of two arbitrary observables of the system. We show that the measurement outcome deviates from the true expectation value of the measured variables unless the coupling strengths between the system and the two detectors are sufficiently weak. This is a striking difference from the case where the commutator of the two observables is a $c$-number. Following the method by Scott and Milburn \cite{A. J. Scott}, we use this generalized model to describe the continuous measurement of two arbitrary observables, and derive both unconditioned and conditioned master equations. We find that the former can be reduced to the Lindblad form in the continuous limit, even if the coupling constant between the system and the apparatus is not infinitesimally small. Finally, using the obtained master equations, we discuss the state preparation of a spin-$1/2$ system by the feedback control based on the simultaneous, continuous measurement of different components of the spin. In our scheme, a target state is obtained as an asymptotic steady state of the time evolution. We find that the effect of measurement and feedback together on the resulting steady state is equivalent to a heat bath, which is similar to the harmonic oscillator case \cite{A. Levy}, and we also derive analytical expressions of the timescale required to reach the steady state. Moreover, the static external magnetic field can generate coherence between the ground state and the excited state. Because of this property, the static external field together with the measurement and feedback control can be utilized to manipulate the state of a TLS in a versatile manner.

This paper is organized as follows. In Sec.~\ref{sec:gakmodel}, we generalize the Arthurs-Kelly measurement model \cite{E. Arthurs, S. L. Braunstein, A. J. Scott} and calculate the average and the variance of the measurement readout. In Sec.~\ref{sec:contmeas}, we simultaneously and continuously measure the system based on this generalized model. The unconditioned and the conditioned master equations are given in this section. In Sec.~\ref{sec:tls}, we perform measurement and feedback control on a TLS in an external magnetic field. Focusing on an asymptotic steady state of the system, we discuss effects of the external field, the measurement, and the feedback control. The summary and conclusion are given in Sec.~\ref{sec:conclusion}.

\section{Measurement model \label{sec:model}}

\subsection{Generalization of Arthurs-Kelly measurement model \label{sec:gakmodel}}

The Arthurs-Kelly model is a single-shot measurement model which can be used to describe the simultaneous measurement of the position and the momentum of a particle in a one-dimensional quantum system \cite{E. Arthurs, S. L. Braunstein, A. J. Scott}. It consists of two detectors and the system to be measured. The pointers of the two detectors are prepared in the Gaussian initial state, and, at an instance of time $t_r$, the position $\hat{x}$ and the momentum $\hat{p}$ of the particle are coupled with the pointers of the two detectors, respectively, when the measurement starts. After the coupling at $t_r$, the positions of the two pointers are influenced by the system; therefore, we can obtain measured values of $\hat{x}$ and $\hat{p}$ from the readouts of the positions of the two pointers by projective measurements of their positions. The coupling between the system and the detectors is described by the time-dependent Hamiltonian $\hat{H}_I(t)$, which is chosen in the following form: 
\begin{align}
\hat{H}_I(t) =  (s_1\hat{x} \hat{p}_1 + s_2 \hat{p} \hat{p}_2)\, \delta(t - t_r),
\end{align}
where $ \hat{p}_i $ $(i = 1,\, 2)$ is the momentum of detector $i$'s pointer, $s_i$ is the coupling strength between the system and detector $i$, and all the quantities here (i.e., $\hat{H}_I$, $s_i$, $\hat{x}$, $\hat{p}$, and $\hat{p}_i$) are dimensionless.

We now consider simultaneous measurements of two arbitrary observables $\hat{A}$ and $\hat{B}$ of the system. This generalized measurement can be performed by the following analogical interaction Hamiltonian between the system and two detectors:
\begin{align}
\hat{H}_I(t) &=  (s_1\hat{A} \hat{p}_1 + s_2 \hat{B} \hat{p}_2)\, \delta(t - t_r).
\end{align} 
Similarly, all the physical quantities discussed here such as $\hat{H}_I$, $s_i$, and $\hat{A}$ are also dimensionless. Moreover, $\hbar$ is set to be unity throughout the whole paper for simplicity \cite{note:units}. Again, detector $i$ is still prepared in the Gaussian initial state $\ket{d_i}$,
\begin{align}
 \left \langle x_i |d_i \right \rangle = (\pi \Delta_i)^{-1/4}\, \exp{\left( -\frac{x_i^2}{2\Delta_i} \right)}, 
\end{align}
where $\ket{x_i}$ is the position eigenstate of detector $i$'s pointer, $\Delta_i \equiv s_i \sigma^2$ $(\textrm{with } s_i > 0)$, and $\sigma^2$ is a parameter characterizing the measurement accuracy.

Before the coupling at $t_r$, the system and the two detectors are assumed to be uncorrelated with each other. Therefore, the density operator of the total system including the system and the detectors initially takes the following form:
\begin{align}
\hat{\rho}_T  \equiv \hat{\rho}_s \otimes |d_1  d_2 \rangle   \langle d_1 d_2 |, 
\label{eq:rhot}
\end{align}  
where $\hat{\rho}_s$ is the partial density operator of the system and $|d_1  d_2 \rangle \equiv |d_1  \rangle \otimes |d_2  \rangle$ is the uncorrelated Gaussian initial state of two detectors. 

After the coupling between the system and the detectors, the total system is in the state
\begin{align}
\hat{\rho}_T^{\prime} = \hat{U}_I \hat{\rho}_T \hat{U}_I^{\dagger},
\end{align}
where 
\begin{align}
\hat{U}_I \equiv \exp\left[ -i (s_1\hat{A} \hat{p}_1 + s_2 \hat{B} \hat{p}_2)\right] 
\end{align}
is the evolution operator during the measurement. As a result, the average of observable $\hat{A}$ and the position of the pointer of detector 1 after the coupling are given as
 \begin{align} 
  \left \langle \hat{A} \right \rangle ^\prime &= \Tr(\hat{A} \hat{\rho}_T^{\prime}) = \Tr(\hat{U}_I^{\dagger} \hat{A} \hat{U}_I \hat{\rho}_T) \equiv \left \langle \hat{U}_I^{\dagger} \hat{A} \hat{U}_I \right \rangle,
  \\
  \left \langle \hat{x}_1 \right \rangle ^\prime &= \Tr(\hat{x}_1 \hat{\rho}_T^{\prime}) = \Tr(\hat{U}_I^{\dagger} \hat{x}_1 \hat{U}_I \hat{\rho}_T) \equiv \left \langle \hat{U}_I^{\dagger} \hat{x}_1 \hat{U}_I \right \rangle.
 \end{align}  
Applying the Baker-Campbell-Hausdorff relation to these two equations, we get
\begin{align}
\left \langle \hat{A} \right \rangle ^\prime = \left \langle \hat{A} \right \rangle - \frac{s_2}{4\sigma^2} \left \langle [\hat{B}, [\hat{B}, \hat{A}]] \right \rangle + O\left(  \frac{s_i ^2}{\sigma^4}\right),
\label{eq: Aaver}
\end{align}
and
\begin{align}
\left \langle \hat{x}_1 \right \rangle ^\prime = s_1 \left[ \left \langle \hat{A} \right \rangle - \frac{s_2}{12\sigma^2} \left \langle [\hat{B}, [\hat{B}, \hat{A}]] \right \rangle + O\left(  \frac{s_i^2}{\sigma^4}\right) \right].
\label{eq:x1aver}
\end{align}   
Here, we only have evenfold commutators since $\left \langle \hat{p}_i^{2n+1} \right \rangle =0$ for the Gaussian state for non-negative integer $n$. Following the same procedure for $\hat{B}$ and $\hat{x}_2$, we can obtain $\left \langle B \right \rangle ^\prime$ and $\left \langle x_2 \right \rangle ^\prime$,
\begin{align}
\left \langle \hat{B} \right \rangle ^\prime =  \left \langle \hat{B} \right \rangle - \frac{s_1}{4\sigma^2} \left \langle [\hat{A}, [\hat{A}, \hat{B}]] \right \rangle + O\left(  \frac{s_i^2}{\sigma^4}\right),
\end{align}
and
\begin{align}
\left \langle \hat{x}_2 \right \rangle ^\prime = s_2 \left[ \left \langle \hat{B} \right \rangle - \frac{s_1}{12\sigma^2} \left \langle [\hat{A}, [\hat{A}, \hat{B}]] \right \rangle + O\left(\frac{s_i^2}{\sigma^4}\right) \right].
\label{eq:x2aver}
\end{align}   

If $\left[ \hat{B}, \left[ \hat{B}, \hat{A} \right] \right]  = \left[ \hat{A}, \left[ \hat{A}, \hat{B} \right] \right]  = 0 $, for instance $\hat{A} = \hat{x}$ and $\hat{B} = \hat{p}$, then all the multifold commutators vanish so that we get $\left \langle \hat{A} \right \rangle ^\prime = s_1^{-1} \left \langle \hat{x}_1 \right \rangle ^\prime $ and $\left \langle \hat{B} \right \rangle ^\prime = s_2^{-1} \left \langle \hat{x}_2 \right \rangle ^\prime $. However, for a general case in which $[\hat{B},[\hat{B},\hat{A}]]$ and $ \left[ \hat{A}, \left[ \hat{A}, \hat{B} \right] \right] $ are nonzero, all the higher-order terms remain and thus the measurement result $ s_i^{-1} \left \langle \hat{x}_i \right \rangle ^\prime $ deviates from $\left \langle \hat{A} \right \rangle ^\prime $ and $\left \langle \hat{B} \right \rangle ^\prime $. From Eqs.~(\ref{eq: Aaver})\,--\,(\ref{eq:x2aver}), we see that the leading order of the deviations is $s_i / \sigma^2$, and thus the deviations are negligible only when $s_i / \sigma^2 \ll 1$. To discuss the deviations, it is convenient to introduce relative deviations $\epsilon_1$ and $\epsilon_2$ defined as
\begin{align}
\epsilon_1 & \equiv \frac{\left \langle \hat{x}_1 \right \rangle ^\prime - s_1 \left \langle \hat{A} \right \rangle ^\prime}{s_1 \left \langle \hat{A} \right \rangle ^\prime},
\label{eq:rd1}\\
\epsilon_2 & \equiv \frac{\left \langle \hat{x}_2 \right \rangle ^\prime - s_2 \left \langle \hat{B} \right \rangle ^\prime}{s_2 \left \langle \hat{B} \right \rangle ^\prime}.
\label{eq:rd2}
\end{align}
For $\hat{A} = \hat{L}_x$ and $\hat{B} = \hat{L}_y$ as an example, where $\hat{L}_x$ and $\hat{L}_y$ are $x$ and $y$ components of the angular momentum, respectively \cite{note:angularmom}, $\epsilon_1$ and $\epsilon_2$, with $ s_i^{-1} \left \langle \hat{x}_i \right \rangle ^\prime $, $\left \langle \hat{A} \right \rangle ^\prime $, and $\left \langle \hat{B} \right \rangle ^\prime $ up to the second order of $s_i / \sigma^2$, read
\begin{align}
\epsilon_1 &= \frac{s_2}{6\sigma^2}\frac{1 - (s_1 + 3s_2)/20\sigma^2}{1 - s_2/4\sigma^2 + s_2(s_1 + 3s_2)/96\sigma^4},
\label{eq:dev1}\\
\epsilon_2 &= \frac{s_1}{6\sigma^2} \frac{1 - (s_2 + 3s_1)/20\sigma^2}{1 - s_1/4\sigma^2 + s_1(s_2 + 3s_1)/96\sigma^4}.
\label{eq:dev2}
\end{align}

\begin{figure}[tb!]
	\centering \includegraphics[scale=0.55]{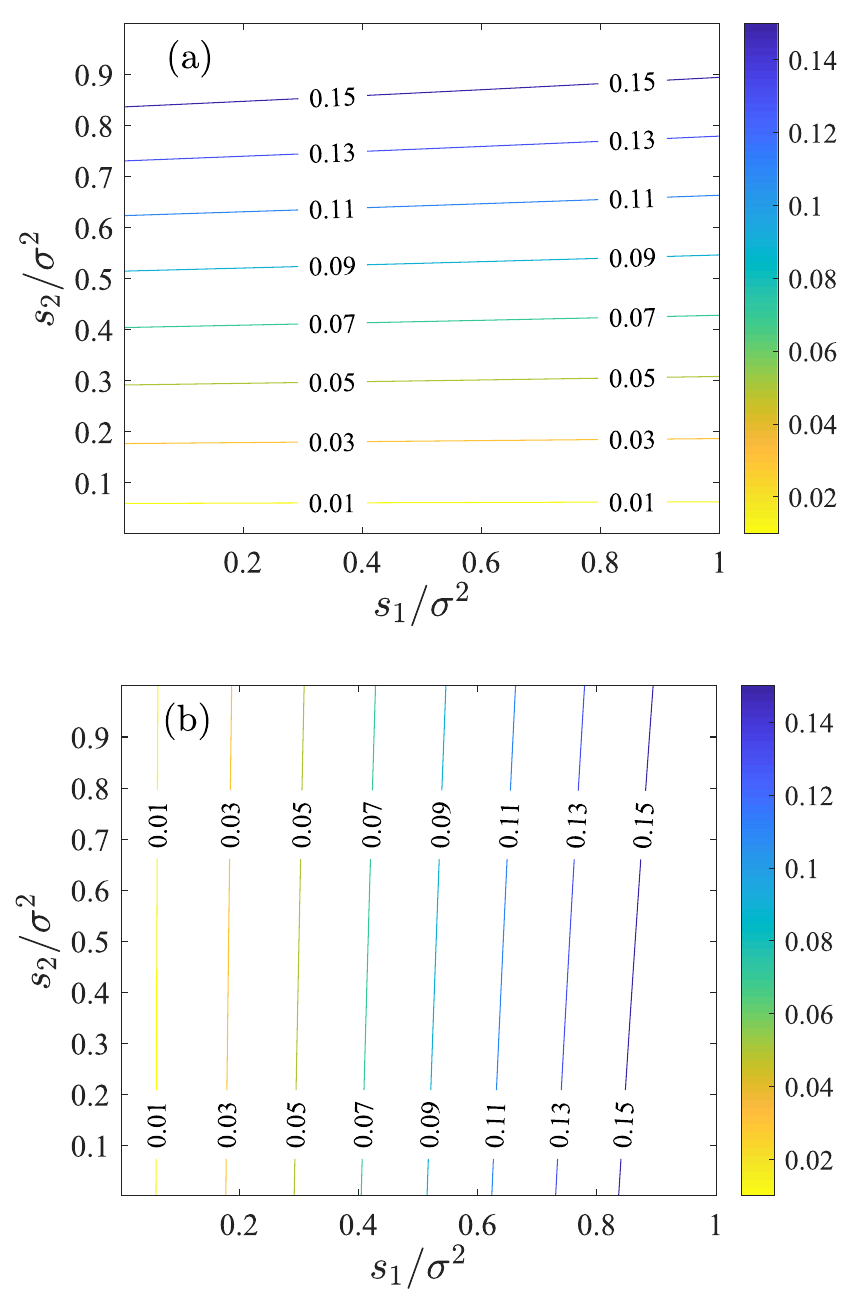}
	\caption{Relative deviations (a) $\epsilon_1$ and (b) $\epsilon_2$ as functions of $s_1 / \sigma^2$ and $s_2 / \sigma^2$ for $\hat{A} = \hat{L}_x$ and $\hat{B} = \hat{L}_y$. $\epsilon_1$ and $\epsilon_2$ are monotonically increasing functions of $s_i/ \sigma^2$.}	
	\label{error}
\end{figure}

In Eqs.~(\ref{eq:dev1}) and~(\ref{eq:dev2}), $\epsilon_1$ and $\epsilon_2$ are independent of the state of the system due to the closed algebra of the angular momentum and the symmetry of the Gaussian state. Take $\epsilon_1$ for example: because of the closed algebra of the angular momentum, the numerator and the denominator of $\epsilon_1$ can be written as a linear combination of $\left\langle \hat{L}_x \right\rangle $, $\left\langle \hat{L}_y \right\rangle $, and $\left\langle \hat{L}_z \right\rangle $. However, due to the symmetry of the Gaussian state, the coefficients of $\left\langle \hat{L}_y \right\rangle $ and $\left\langle \hat{L}_z \right\rangle $, which are averages of odd powers of $\hat{x}_i$ and $\hat{p}_i$, are zero. Therefore, the numerator and the denominator of $\epsilon_1$ are proportional to $\left\langle \hat{L}_x \right\rangle $, which are canceled with each other finally. The relative deviations $\epsilon_1$ and $\epsilon_2$ are monotonically increasing with parameters $s_1 / \sigma^2$ and $s_2 / \sigma^2$. When $s_i / \sigma^2 \simeq 0.5$, the relative deviations $\epsilon_1$ and $\epsilon_2$ reach around $10\%$, which are non-negligible. Consequently, $s_i / \sigma^2 $ must be much smaller than 1 in order to obtain an accurate measurement outcome. Details are shown in Fig.~\ref{error}.

We now go back to the arbitrary observables $\hat{A}$ and $\hat{B}$, and consider the second moment of the positions of the two pointers $\hat{x}_1$ and $\hat{x}_2$ after the coupling, 
\begin{align}
\left \langle \hat{x}_1^2 \right \rangle ^\prime = & \frac{s_1\sigma^2}{2} \left[ 1 + \frac{2s_1}{\sigma^2}\left \langle \hat{A}^2 \right \rangle - \frac{s_1 s_2}{12 \sigma^4} \Big(  \left \langle [\hat{B},[\hat{B},\hat{A}^2]] \right \rangle \right.
\notag
\\
 & \left. + \left\langle \hat{A}\, [\hat{B},[\hat{B}, \hat{A}]] \right\rangle  + 
\left\langle [\hat{B},\hat{A}\, [\hat{B}, \hat{A}]] \right\rangle \Big)  + O \left( \frac{s_i^3}{\sigma^6} \right) \right],
\label{eq:x1var}
\end{align}
and
\begin{align}
\left \langle \hat{x}_2^2 \right \rangle ^\prime = & \frac{ s_2\sigma^2}{2} \left[ 1 + \frac{2s_2}{\sigma^2} \left \langle \hat{B}^2 \right \rangle - \frac{s_1s_2}{12 \sigma^4}\Big( \left \langle [\hat{A},[\hat{A},\hat{B}^2]] \right \rangle \right.
\notag
\\
& \left. + \left\langle \hat{B}\, [\hat{A},[\hat{A}, \hat{B}]] \right\rangle + 
\left\langle [\hat{A},\hat{B}\, [\hat{A}, \hat{B}]] \right\rangle  \Big) + O\left( \frac{s_i^3}{\sigma^6}\right) \right].
\label{eq:x2var}
\end{align}
The leading term of the variance $\left \langle \hat{x}_i^2 \right \rangle ^\prime - \left( \left \langle \hat{x}_i \right \rangle ^\prime\right) ^2$ of the measurement result depends on the parameter $s_i \sigma^2$. In order to get a stable readout, i.e., the variances are small so that the measurement results are less scattered, $s_i \sigma^2$ need to be set as small as possible. There is a tradeoff between the stability characterized by $s_i \sigma^2$ and the accuracy characterized by $s_i/ \sigma^2$ [see Eqs.~(\ref{eq:dev1}) and (\ref{eq:dev2})]. Therefore, $\sigma^2$ cannot be too large for a given $s_i$, although $\sigma^2$ is intrinsically a large quantity in the weak measurements, which is assumed to hold $s_i/\sigma^2 \ll 1$. On the other hand, for a given $\sigma^2$, $s_i$ should be sufficiently small so that both $s_i \sigma^2$ and $s_i/ \sigma^2$ are smaller than unity. From this point of view, the generalized Arthurs-Kelly measurement model is valid only for a weak measurement.

\subsection{Continuous measurement \label{sec:contmeas}}
A continuous and simultaneous measurement of observables $\hat{A}$ and $\hat{B}$ can be realized by the following interaction Hamiltonian \cite{A. J. Scott}:
\begin{align}
 \hat{H}_I(t) = \sum_{n=1}^{\infty} (s_1\hat{A} \hat{p}_1 + s_2\hat{B} \hat{p}_2)\, \delta(t - n\delta t),
\end{align} 
where $\delta t$ is the time interval between two consecutive measurements and will finally be set to be infinitesimal. After each single-shot measurement, the readouts of $x_1$ and $x_2$ are recorded, and then the total composite system is reset to the decoupled initial state given by Eq.~(\ref{eq:rhot}) based on the Born-Markov assumption. The state of the system immediately before and after the $n$th measurement is denoted by $\hat{\rho}_s(n\delta t)$ and $\hat{\rho}_s^\prime(n\delta t)$ respectively, which satisfy the following relationship:
\begin{align}
\hat{\rho}^\prime_s(n\delta t) =  \Tr _d \left[ \hat{U}_I \hat{\rho}_s(n\delta t)  \otimes |d_1  d_2 \rangle  \langle d_1 d_2 |   \hat{U}_I^\dagger\right] , 
\label{eq:rhos}
\end{align}
where $\Tr_d $ means the trace over the degrees of freedom of the two detectors. 

We introduce the following measurement operator: 
\begin{align}
\hat{M}(x_1,x_2)  \equiv & \left\langle x_1 x_2| \hat{U}_I |d_1 d_2\right\rangle 
\notag
\\
 = & \int dp_1 dp_2\, \left\langle  x_1 x_2| \hat{U}_I |p_1 p_2 \right\rangle \left\langle p_1p_2|d_1 d_2\right\rangle 
\notag
\\
 = & (2\pi)^{-1} \int dp_1 dp_2\ \exp \left\lbrace  i [(x_1 - s_1\hat{A})p_1 \right.
\notag
\\
& \left. + (x_2 -s_2\hat{B})p_2] \right\rbrace  \left\langle p_1p_2|d_1 d_2\right\rangle, 
\end{align}
where $\ket{x_1 x_2} \equiv \ket{x_1} \otimes \ket{x_2}$, $\ket{p_1 p_2} \equiv \ket{p_1} \otimes \ket{p_2}$, and $\ket{p_i}$ is the eigenstate of the momentum of detector $i$'s pointer. Equation~(\ref{eq:rhos}) can be rewritten explicitly as
\begin{align}
\hat{\rho}^\prime_s(n\delta t) = & \int dx_1 dx_2 \ \hat{M}(x_1,x_2)\, \hat{\rho}_s(n\delta t)\, \hat{M}^\dagger(x_1,x_2)
\notag
\\
= & \int dp_1 dp_2 \ \exp \left[ -i (s_1p_1 \hat{A} + s_2 p_2 \hat{B})\right] \, \hat{\rho}_s(n\delta t)  
\notag
\\
& \exp\left[ i (s_1p_1 \hat{A} + s_2 p_2 \hat{B})\right]  \, \left| \left\langle p_1p_2|d_1d_2\right\rangle  \right| ^2
\notag
\\
= & \hat{\rho}_s(n\delta t) - \frac{s_1}{4 \sigma^2} [\hat{A},[\hat{A},\hat{\rho}_s(n\delta t)]] 
\notag
\\
 &- \frac{s_2}{4 \sigma^2} [\hat{B},[\hat{B},\hat{\rho}_s(n\delta t)]] + O\left( \frac{s_i^2}{\sigma^4}\right).  \label{eq:mast}
\end{align}
Note that the right-hand side of the last equality of Eq.~(\ref{eq:mast}) contains neither the terms $[ \hat{A}, [\hat{B}, \hat{\rho}_s]]$ nor $[ \hat{B}, [\hat{A}, \hat{\rho}_s]]$, which describe the interplay effect between the measurements of the two observables, because their coefficients are zero for Gaussian states. 

Taking the unitary evolution by the system Hamiltonian $\hat{H}_s$ after each measurement into consideration, the state of the system becomes
\begin{align}
\hat{\rho}_s(n\delta t + \delta t) = \hat{U}_s\, \hat{\rho}^\prime_s(n\delta t)\,
\hat{U}_s^{\dagger},
\end{align}
where
\begin{align}
\hat{U}_s \equiv \exp \left( -i \hat{H}_s \delta t \right)  
\end{align}
is the unitary evolution operator between two consecutive measurements. After taking the continuous limit $\delta t \to 0$ and $\sigma \to \infty$ with $\zeta \equiv 1/(\delta t\, \sigma^2) = \mathrm{const.}$, the unconditioned master equation of the system under the simultaneous and continuous measurement reduces to the Lindblad form \cite{H.-P. Breuer, H. M. W., K. Jacobs2, Lindblad, Gorini76}:
\begin{align}
\frac{d\hat{\rho}_s}{dt} = -i[\hat{H}_s, \hat{\rho}_s] - \frac{\gamma_1}{8}[\hat{A}, [\hat{A}, \hat{\rho}_s]] - \frac{\gamma_2}{8} [\hat{B}, [\hat{B}, \hat{\rho}_s]].
\label{eq:master}
\end{align}
Here, $\gamma_i \equiv 2s_i \zeta$ is the measurement strength of $\hat{A}$ $(i = 1)$ or $\hat{B}$ $(i = 2)$. In addition, we can obtain the master equation for the measurement of a single observable $\hat{A}$ or $\hat{B}$ by setting $s_2$ or $s_1$ to be zero, respectively. Note that even for noninfinitesimal $s_i$, the master equation (\ref{eq:master}) for simultaneous and continuous measurement is still valid in the continuous limit $s_i / \sigma^2 \to 0$, unlike the results for the single-shot measurement in the previous section. 

The final master equation (\ref{eq:master}) of simultaneous measurement does not contain terms of the interplay effect such as $[ \hat{A}, [\hat{B}, \hat{\rho}_s]]$ and $[ \hat{B}, [\hat{A}, \hat{\rho}_s]]$, but just consists of a linear combination of the two independent measurement effects. This is because of the uncorrelated Gaussian initial state and the Born-Markovian approximation: Before each single-shot measurement, the total composite system is reset to the decoupled initial state given by Eq.~(\ref{eq:rhot}). In addition, according to Eq.~(\ref{eq:mast}), the single-shot measurement discussed in the previous section does not introduce the terms of the interplay effect in the master equation if the initial state of the system and the two detectors are uncorrelated with each other and the detectors are initially prepared in the Gaussian state. Therefore, the terms of the interplay effects $[ \hat{A}, [\hat{B}, \hat{\rho}_s]]$ and $[ \hat{B}, [\hat{A}, \hat{\rho}_s]]$ are absent in the final master equation. Note that for $\hat{A} = \hat{x}$ and $\hat{B} = \hat{p}$ considered in Ref.~\cite{A. J. Scott}, such terms of the interplay effect have vanished, $[\hat{x}, [\hat{p}, \hat{\rho}_s]] - [ \hat{p}, [\hat{x}, \hat{\rho}_s]] = 0$, because of $[\hat{x} , \hat{p}] = i$. Thus, even though the master equation of the simultaneous and continuous measurement of $\hat{x}$ and $\hat{p}$ does not contain the terms of the interplay effect of the two measurements \cite{A. J. Scott}, it is still open whether this equation still holds for arbitrary observables $\hat{A}$ and $\hat{B}$. Here, we have shown that the master equation indeed does not contain the terms of the interplay effect, irrespective of the observables measured. 

Moreover, we shall also derive the conditioned master equation of simultaneous measurement. The $n$th measurement outcome of detector $i$ is denoted by $x_{i}(n)$ and the final master equation is conditioned by a sequence of the outcome $\{x_{i}(n)\}$. The average and the variance of $x_{i}(n)$ can be obtained by keeping the leading terms of Eqs.~(\ref{eq:x1aver}),~(\ref{eq:x2aver}),~(\ref{eq:x1var}), and~(\ref{eq:x2var}),
\begin{align}
E\left[  x_{1}(n)\right]   & \approx s_1 \left\langle \hat{A}\right\rangle, 
\\
E\left[  x_{2}(n)\right]   & \approx s_2 \left\langle \hat{B}\right\rangle, 
\\
V\left[  x_{i}(n)\right]   & =  \left \langle \hat{x}_i^2 \right \rangle ^\prime - \left( \left \langle \hat{x}_i\right \rangle ^\prime \right) ^2 
\approx \frac{s_i }{2\zeta \delta t}, \label{eq:var}
\end{align}
where $E[\cdots]$ represents the ensemble average over all the experimental realizations and $V[\cdots]$ represents the variance of the measurement outcome. The correlation function of $\hat{x}_1$ and $\hat{x}_2$ after the coupling can also be obtained:
\begin{align}
\left\langle \hat{x}_1 \hat{x}_2\right\rangle ^{\prime} = & \Tr\left( \hat{x}_1 \hat{x}_2 \hat{\rho}_T^{\prime}\right) 
\notag
\\
= & \frac{s_1 s_2}{2} \left\langle \hat{A} \hat{B} + \hat{B} \hat{A} \right\rangle. 
\label{eq:corr}
\end{align}
Comparing Eqs.~(\ref{eq:var}) and~(\ref{eq:corr}), we find the covariance $C[x_1(n), x_2(n)] \equiv \avg{\hat{x}_1 \hat{x}_2}' - \avg{\hat{x}_1}' \avg{\hat{x}_2}'$ of $\hat{x}_1$ and $\hat{x}_2$ [which is of the order of $(\delta t)^0$] is much smaller than the variance of $\hat{x}_1$ and $\hat{x}_2$ (which is of the order of $1/\delta t$) in the continuous limit $\delta t \to 0$. Therefore, the correlation between $\hat{x}_1$ and $\hat{x}_2$ can be neglected within the current approximation where the terms of order $1/\delta t$ are kept for the second moment of $x_i$, and $x_{1}(n)$ and $x_{2}(n)$ can be treated as two uncorrelated random variables \cite{note:covar}. Then, approximating $x_{i}(n)$ as a Gaussian random variable, $x_{i}(n)$ can be written as a summation of its average and fluctuation,
\begin{align}
x_{1}(n) &= s_1 \left\langle \hat{A}\right\rangle + \sqrt{\frac{s_1}{2\zeta}}\, d\xi_1(n) \cdot (\delta t)^{-1}
\notag
\\ 
&\equiv s_1 \left\langle \hat{A}\right\rangle + \lambda_1\, d\xi_1(n) \cdot (\delta t)^{-1},  \label{eq:out1}
\end{align}
and
\begin{align}
x_{2}(n) &= s_2 \left\langle \hat{B}\right\rangle + \sqrt{\frac{s_2}{2\zeta}}\,
d\xi_2(n) \cdot (\delta t)^{-1}
\notag
\\ 
&\equiv s_2 \left\langle \hat{B}\right\rangle + \lambda_2\, d\xi_2(n) \cdot (\delta t)^{-1},   \label{eq:out2}
\end{align}
where $\lambda_i \equiv \sqrt{ s_i / 2\zeta } $ is the  fluctuation of the measurement outcome of observables $\hat{A}$ ($i = 1$) and $\hat{B}$ ($i = 2$), and $d\xi_i$ is the It\^{o} increment \cite{C. W., N. G. V. Kampen} which satisfies 
\begin{align}
& E[d\xi_i] = 0, 
\\
& E[d\xi_i(m) \cdot d\xi_i(n)] = \delta_{mn} \cdot \delta t,
\\
& E[d\xi_1(n) \cdot d\xi_2(n)] = 0.
\end{align}
For notational simplicity, we will treat $d\xi_i$ to be equal to $\delta t^{1/2}$ and omit the symbol $E[\cdots]$ in the following derivation. Then, the measurement operator can be expanded as
\begin{widetext}
\begin{align}
\hat{M}(x_1(n),x_2(n))  =&  (2\pi )^{-1} \int dp_1 dp_2 \ 
\exp \left\lbrace  i[x_1(n) p_1 + x_2(n) p_2 ] \right\rbrace 
\exp \left[   -i (s_1 p_1\hat{A} + s_2 p_2\hat{B}) \right] 
\notag
\\
&\cdot \frac{(\Delta_1 \Delta_2)^{-1/4}}{\pi^{1/2}}
\exp \left(  -\frac{\Delta_1 p_1^2 + \Delta_2 p_2^2}{2}  \right) 
\notag
\\
\propto & \int dp_1 dp_2 \ \exp \left\lbrace  -i (s_1 p_1\hat{A} 
+ s_2 p_2\hat{B} ) \right\rbrace  
\exp  \left[  -\frac{\Delta_1 \left(  p_1 - \frac{i x_1(n)}{\Delta_1} \right) ^2 + \Delta_2 \left(  p_2 - \frac{i x_2(n)}{\Delta_2} \right) ^2}{2} \right]  
\notag
\\
= & \int dp_1 dp_2 \ \left\lbrace \sum_{n = 0}^{\infty}\frac{1}{n!}  \left[  -i (s_1 p_1 \hat{A} + s_2 p_2 \hat{B} ) \right]  ^n \right\rbrace \cdot
\exp \left[   -\frac{\Delta_1 \left(  p_1 - \frac{i x_1(n)}{\Delta_1} \right) ^2 + \Delta_2 \left(  p_2 - \frac{i x_2(n)}{\Delta_2} \right)  ^2}{2}  \right] 
\notag
\\
= & \ \frac{2\pi }{(\Delta_1 \Delta_2)^{1/2}} \bigg[ 1 + 
\frac{s_1 \hat{A}}{\Delta_1}x_1(n) + \frac{s_2 \hat{B}}{\Delta_2}x_2(n)
- \frac{s_1^2 \hat{A}^2 }{2\Delta_1}
+ \frac{s_1^2 \hat{A}^2 }{2\Delta_1^2}x_1(n)^2 - \frac{s_2^2 \hat{B}^2 }{2\Delta_2} + \frac{s_2^2 \hat{B}^2 }{2\Delta_2^2}x_2(n)^2 \bigg]  + O\left( \delta t ^{3/2}\right)
\notag
\\ 
\propto &\  1 + \zeta \delta t \left(  s_1 \left\langle 
\hat{A}\right\rangle \hat{A} + s_2 \left\langle 
\hat{B}\right\rangle \hat{B} - \frac{s_1 \hat{A}^2}{4}- \frac{s_2 \hat{B}^2}{4} \right) 
+ \lambda_1 \zeta\, \hat{A}\, d\xi_1 + \lambda_2 \zeta\, \hat{B}\, d\xi_2 + O\left( \delta t ^{3/2}\right).
\end{align}
\end{widetext}
The unnormalized state of the system after monitoring then becomes
\begin{widetext}
\begin{align}
\hat{\rho}^\prime_s(n\delta t) = & \hat{M}\, \hat{\rho}_s(n\delta t)\, \hat{M}^{\dagger}
\notag
\\ 
= & \hat{\rho}_s + \zeta\, \delta t \bigg[  s_1 \left\langle 
\hat{A}\right\rangle (\hat{A } \hat{\rho}_s + \hat{\rho}_s \hat{A } ) +
s_2 \left\langle \hat{B}\right\rangle (\hat{B} \hat{\rho}_s
+ \hat{\rho}_s \hat{B} ) - \frac{s_1}{4} (\hat{A}^2
\hat{\rho}_s + \hat{\rho}_s \hat{A}^2 ) - \frac{s_2}{4} (\hat{B}^2
\hat{\rho}_s + \hat{\rho}_s \hat{B}^2 )\bigg]
\notag
\\ 
& + \lambda_1\, \zeta\, (\hat{A} \hat{\rho}_s + \hat{\rho}_s \hat{A})\, d\xi_1
+ \lambda_2\, \zeta\, (\hat{B} \hat{\rho}_s + \hat{\rho}_s \hat{B})\, d\xi_2
+  \zeta^2\, \delta t\, (\lambda_1^2\, \hat{A}\, \hat{\rho}_s\, \hat{A} + \lambda_2^2\, \hat{B}\, \hat{\rho}_s\, \hat{B}) + O\left( \delta t ^{3/2}\right). 
\end{align}
\end{widetext}
The normalization constant for the state after the measurement reads
\begin{align}
\Tr (\hat{M}\, \hat{\rho}_s\, \hat{M}^{\dagger}) = & 1 +  2 \zeta\, \delta t\, \left( s_1 \left\langle \hat{A}\right\rangle ^2 + s_2 \left\langle  \hat{B}\right\rangle ^2\right) 
\notag
\\ 
& + 2\lambda_1\, \zeta\, \left\langle \hat{A}\right\rangle\, d\xi_1   +  2\lambda_2\, \zeta\, \left\langle \hat{B}\right\rangle\, d\xi_2 + O\left( \delta t ^{3/2}\right).   
\end{align}
Applying $(1 + x)^{-1} \approx 1 - x + x^2$ and keeping terms up to the first order of $\delta t$, we obtain the normalized state after the measurement,
\begin{widetext}
\begin{align}
\hat{\rho}^\prime_s(n\delta t) = & 
\frac{\hat{M}\hat{\rho}_s \hat{M}^{\dagger}}{\Tr (\hat{M}\hat{\rho}_s \hat{M}^{\dagger})} 
\notag
\\ 
\approx & \hat{\rho}_s - \frac{s_1 \zeta}{4}(\hat{A}^2 \hat{\rho}_s +
\hat{\rho}_s \hat{A}^2 - 2 \hat{A} \hat{\rho}_s \hat{A})\, \delta t 
- \frac{s_2 \zeta}{4} (\hat{B}^2 \hat{\rho}_s + \hat{\rho}_s \hat{B}^2 - 2 \hat{B} \hat{\rho}_s\hat{B})\, \delta t + \sqrt{2s_1 \zeta}\, \bigg( 
\frac{\hat{A}\hat{\rho}_s + \hat{\rho}_s \hat{A}}{2} 
 - \left\langle \hat{A}\right\rangle \hat{\rho}_s \bigg) \, d\xi_1
\notag
\\
& + \sqrt{2s_2 \zeta}\, \bigg( 
\frac{\hat{B}\hat{\rho}_s + \hat{\rho}_s\hat{B}}{2} - \left\langle \hat{B}\right\rangle \hat{\rho}_s \bigg) \, d\xi_2
\notag
\\ 
= & \hat{\rho}_s -\frac{\gamma_1}{8}[\hat{A}, [\hat{A}, \hat{\rho}_s]]\, \delta t - \frac{\gamma_2}{8}[\hat{B}, [\hat{B}, \hat{\rho}_s]]\, \delta t + \sqrt{\gamma_1}\, \mathcal{H}  \left[  \left( \hat{A} - \left \langle \hat{A} \right \rangle \right)  \, \hat{\rho}_s \right] \, d\xi_1 + \sqrt{\gamma_2}\, \mathcal{H}  \left[  \left( \hat{B} - \left \langle \hat{B} \right \rangle \right) \, \hat{\rho}_s \right] \, d\xi_2. 
\end{align}
\end{widetext}
Here, the symbol $\mathcal{H}[\hat{O}]$ is defined as the Hermitian part of operator $\hat{O}$,
\begin{align}
\mathcal{H}[\hat{O}]  \equiv \frac{1}{2}(\hat{O} + \hat{O}^\dagger).
\end{align}
Including the unitary evolution between the two consecutive measurements and taking the continuous limit, the conditioned master equation reads
\begin{align}
d\hat{\rho}_s= & -i[\hat{H}_s, \hat{\rho}_s]\, dt -\frac{\gamma_1}{8}[\hat{A}, [\hat{A}, \hat{\rho}_s]]\, dt - \frac{\gamma_2}{8}[\hat{B}, [\hat{B}, \hat{\rho}_s]]\, dt 
\notag
\\
& +\sqrt{\gamma_1}\, \mathcal{H} \left[  \left(  \hat{A} - \left \langle \hat{A} \right \rangle  \right) \hat{\rho}_s \right] \, d\xi_1 
\notag
\\
& + \sqrt{\gamma_2}\, \mathcal{H} \left[ \left( \hat{B} - \left\langle  \hat{B} \right\rangle  \right) \hat{\rho}_s \right] \, d\xi_2.
\label{eq:cme}
\end{align}

\section{state preparation by the feedback control \label{sec:tls}} 

We now propose a scheme to manipulate the state of a spin-$1/2$ system based on the generalized Arthurs-Kelly measurement model obtained in the previous section. For simultaneous and continuous measurement of the $x$ and $y$ components of the spin $\boldsymbol{\hat{S}}$, i.e., $\hat{A} = \hat{\sigma}_x /2$ and $\hat{B} = \hat{\sigma}_y /2$ with $\hat{\sigma}_x$ and $\hat{\sigma}_y$ being the Pauli matrices, the conditioned master equation (\ref{eq:cme}) can be written as
\begin{align}
 d\hat{\rho}_s= & -i[\hat{H}_s, \hat{\rho}_s] dt -\frac{\Gamma_x}{8}[\hat{\sigma}_x, [\hat{\sigma}_x, \hat{\rho}_s]] dt - \frac{\Gamma_y}{8}[\hat{\sigma}_y, [\hat{\sigma}_y, \hat{\rho}_s]] dt 
\notag
\\
& +\sqrt{\Gamma_x} \mathcal{H} \left[  (\hat{\sigma}_x - \left \langle \hat{\sigma}_x \right \rangle) \,\hat{\rho}_s\right] \, d\xi_x
\notag
\\
& + \sqrt{\Gamma_y} \mathcal{H} \left[ (\hat{\sigma}_y - \left \langle \hat{\sigma}_y \right \rangle) \,\hat{\rho}_s \right] \, d\xi_y, 
\end{align}
where $\Gamma_x \equiv \gamma_1/4 = s_1\zeta/2$ and $\Gamma_y \equiv \gamma_2/4= s_2\zeta/2$ are the measurement strengths of the $x$ and $y$ components of the spin, respectively. We assume that the system is in a static external magnetic field; as a consequence, the Hamiltonian of the system can be represented as 
\begin{align}
\hat{H}_s = \omega_x \hat{\sigma_x} + \omega_y \hat{\sigma}_y + \omega_z \hat{\sigma}_z, 
\label{eq:Hs}
\end{align}
where $\omega_x $, $\omega_y $, and $\omega_z $ can be set by the magnitude and the direction of the static external field.

The measurement signals $\bar{\sigma}_x$ and $\bar{\sigma}_y$ are defined as \cite{A. Levy}
\begin{align}
\bar{\sigma}_xdt &= \left \langle \hat{\sigma}_x \right \rangle dt + \frac{d\xi_x}{\sqrt{\Gamma_x}}, \label{eq: sigmax}\\
\bar{\sigma}_ydt &= \left \langle \hat{\sigma}_y \right \rangle dt + \frac{d\xi_y}{\sqrt{\Gamma_y}}, \label{eq: sigmay}
\end{align}
which will be fed back to the system without time delay \cite{T. L. Patti} to control an additional external magnetic field $\boldsymbol{B}$ by the following feedback control Hamiltonian:
\begin{align}
\hat{H}_f\, dt  =&  \boldsymbol{B} \cdot  \boldsymbol{\hat{S}}\, dt
\notag
\\
 =& (\alpha_1 \bar{c} + \alpha_1^* \bar{c}^*)\hat{\sigma}_x\, dt +  (\alpha_2 \bar{c} + \alpha_2^* \bar{c}^*)\hat{\sigma}_y\, dt
\notag
\\
& +  (\alpha_3 \bar{c} + \alpha_3^* \bar{c}^*)\hat{\sigma}_z\, dt
\notag
\\
= &  \bar{c}dt(\alpha_1 \hat{\sigma}_x + \alpha_2 \hat{\sigma}_y + \alpha_3 \hat{\sigma}_z) + \textrm{H.c.},
\label{eq:fb}
\end{align}
where $\alpha_i$ ($i = x, y, z$) is an arbitrary complex number used to control the  $i$ component of the additional magnetic field $\boldsymbol{B}$, $\bar{c}dt \equiv\frac{1}{2}(\bar{\sigma}_x - i\bar{\sigma}_y)dt$ is the complex measurement signal, and $\textrm{H.c.}$ is the Hermitian conjugate of the former term. Here, we have set the Lande factor and the Bohr magneton to be unity for simplicity. Equation~(\ref{eq:fb}) can be rewritten into an equivalent, but more compact form,
\begin{align}
\hat{H}_f\, dt = -i \kappa_f\, \bar{c}\, dt \cdot \hat{F} + \textrm{H.c.},
\end{align}
where $\kappa_f $ is an arbitrary real positive parameter called feedback control strength, and $\hat{F}$ is a linear combination of $\hat{\sigma}_x$, $\hat{\sigma}_y$, and $\hat{\sigma}_z$, corresponding to $\hat{F} = i\kappa_f^{-1} (\alpha_1 \hat{\sigma}_x + \alpha_2 \hat{\sigma}_y + \alpha_3 \hat{\sigma}_z)$ in terms of $\alpha_i$ in Eq.~(\ref{eq:fb}).

Suppose the system is in the state $\hat{\rho}_s$ at time $t$; the state of the system at $t + dt$ after the simultaneous, continuous measurement and feedback control is given by $\exp(-i\hat{H}_f dt )\, (\hat{\rho}_s + d\hat{\rho}_s)\, \exp(i\hat{H}_f dt  )$. Within the first order in $dt$, we need to keep the following terms: $\hat{\rho}_s + d\hat{\rho}_s$, $-i[\hat{H}_f dt, \hat{\rho}_s]$, $-i [\hat{H}_f dt, d\hat{\rho}_s]$, and $-2^{-1}[\hat{H}_f dt,[\hat{H}_f dt, \hat{\rho}_s]]$. Note that the last one also has the first order of $dt$ due to the It\^{o} rule. By inserting Eqs.~(\ref{eq: sigmax}) and~(\ref{eq: sigmay}) into these terms, we obtain the ensemble-averaged master equation of the system under the simultaneous, continuous measurement and the feedback control,
\begin{align}
\frac{d\hat{\rho}_s}{dt} =& -i[\hat{H}_s, \hat{\rho}_s] + \frac{\Gamma_y}{2}\mathcal{D}[\hat{c}]\hat{\rho}_s + \frac{\Gamma_y}{2}\mathcal{D}[\hat{c}^\dagger]\hat{\rho}_s 
\notag
\\
& + \frac{\Gamma_x - \Gamma_y}{4}\mathcal{D}[\hat{c} + \hat{c}^\dagger]\hat{\rho}_s +\frac{\kappa_f^2}{4\Gamma_x}\mathcal{D}[i(\hat{F} - \hat{F}^\dagger)]\hat{\rho}_s  
\notag
\\
& +\frac{\kappa_f^2}{4\Gamma_y}\mathcal{D}[\hat{F} + \hat{F}^\dagger]\hat{\rho}_s - \frac{\kappa_f}{2} \big([\hat{F},\hat{c}\hat{\rho}_s]-[\hat{F}^\dagger,\hat{\rho}_s \hat{c}^\dagger]
\notag
\\
& +[\hat{F},\hat{\rho}_s \hat{c}] - [\hat{F}^\dagger, \hat{c}^\dagger\hat{\rho}_s] \big),  \label{eq:GME}
\end{align} 
where $\hat{c} \equiv \frac{1}{2} (\hat{\sigma}_x - i \hat{\sigma}_y)$ is the lowering operator, and the superoperator $\mathcal{D}[\hat{O}]$ is defined for an arbitrary operator $\hat{O}$ by 
\begin{align}
\mathcal{D}[\hat{O}] \, \hat{\rho}_s \equiv \hat{O} \hat{\rho}_s  \hat{O}^{\dagger} - \frac{1}{2}\left(\hat{O}^{\dagger} \hat{O} \hat{\rho}_s   +  \hat{\rho}_s  \hat{O}^{\dagger} \hat{O}\right).
\end{align}	

Equation~(\ref{eq:GME}) provides the general form of the evolution of the system. The form of the operator $\hat{F}$ is to be chosen according to the target state. Note that, for $\hat{F} = \hat{c}^{\dagger}$ [i.e., $\alpha_1 = - i\kappa_f /2$, $\alpha_2 = \kappa_f /2$, and $\alpha_3 = 0$ in Eq.~(\ref{eq:fb})], Eq.~(\ref{eq:GME}) can be written in the Lindblad form,
\begin{align}
\frac{d\hat{\rho}_s}{dt} = -i[\hat{H}_s , \hat{\rho}_s] + k_1\mathcal{D}[\hat{c}]\hat{\rho}_s + k_2\mathcal{D}[\hat{c}^\dagger]\hat{\rho}_s + k_3\mathcal{D}[\hat{c} + \hat{c}^\dagger]\hat{\rho}_s,
\label{eq:ME}
\end{align} 
with
\begin{align}
k_1 & \equiv \frac{\Gamma_y}{2} + \frac{\kappa_f^2}{2\Gamma_x} + \kappa_f, \label{eq:k1}
\\
k_2 & \equiv \frac{\Gamma_y}{2} + \frac{\kappa_f^2}{2\Gamma_x} - \kappa_f, \label{eq:k2}
\\
k_3 & \equiv \frac{\Gamma_x - \Gamma_y}{4} - \frac{\kappa_f^2}{4\Gamma_x} + \frac{\kappa_f^2}{4\Gamma_y}. \label{eq:k3}
\end{align}
The first term in the right-hand side of Eq.~(\ref{eq:ME}) represents the unitary evolution governed by the static external field while the latter three terms represent the effect of the simultaneous measurement and the feedback control. 

For the spin-$1/2$ system, it is more convenient to discuss the problem in the Bloch coordinate system and represent $\hat{\rho}_s$ with the basis $\hat{\sigma}_x$, $\hat{\sigma}_y$, $\hat{\sigma}_z$, and the unit operator $\hat{I}$, 
\begin{align}
\hat{\rho}_s = \frac{1}{2}(x \hat{\sigma}_x + y \hat{\sigma}_y + z \hat{\sigma}_z + \hat{I}),
\label{eq:rho_s}
\end{align}
where $(x, y, z) \in \mathbb{R}^3$ is the Bloch vector satisfying $\sqrt{x^2 + y^2 + z^2} \leqslant 1$. By inserting Eq.~(\ref{eq:rho_s}) into Eq.~(\ref{eq:ME}) and using $\Tr(\hat{\sigma}_j \hat{\rho}_s) = j$\, (with $j = x, y, z$), we obtain
\begin{align}
\dot{x} &= 2(\omega_y z - \omega_z y) -\frac{k_1 + k_2 }{2} x, \label{eq:xdot}
\\
\dot{y} &= 2(\omega_z x - \omega_x z) -\frac{k_1 + k_2 + 4k_3}{2} y, \label{eq:ydot}
\\
\dot{z} &= 2(\omega_x y - \omega_y x) -(k_1 + k_2 + 2k_3)z + k_2 - k_1. \label{eq:zdot}
\end{align} 
The steady solution of these master equations is obtained by setting $\dot{x} = \dot{y} = \dot{z} = 0$ in Eqs.~(\ref{eq:xdot})\,--\,(\ref{eq:zdot}) \cite{note:explain},
\begin{align}
x_s &= 4\eta ^{-1} (k_2 - k_1)[\omega_y (k_1 + k_2 + 4k_3) + 4\omega_x \omega_z ],
\label{eq:xs} 
\\
y_s &= 4\eta ^{-1} (k_1 - k_2)[\omega_x (k_1 + k_2 ) - 4\omega_y \omega_z ],
\label{eq:ys} 
\\
z_s &= \eta ^{-1} (k_2 - k_1) \left[ (k_1 + k_2) (k_1 + k_2 + 4k_3) 
+ 16\omega_z^2\right],  
\label{eq:zs}
\end{align} 
with
\begin{align}
\eta \equiv & (k_1 + k_2) (k_1 + k_2 + 2k_3)(k_1 + k_2 + 4k_3) + 8\omega_x ^2 (k_1 + k_2)
\notag
\\
&+ 8\omega_y ^2 (k_1 + k_2 + 4k_3) + 16\omega_z^2(k_1 + k_2 + 2k_3). 
\label{eq:eta}
\end{align}

\begin{figure}[tb!]
	\centering \includegraphics[scale=0.55]{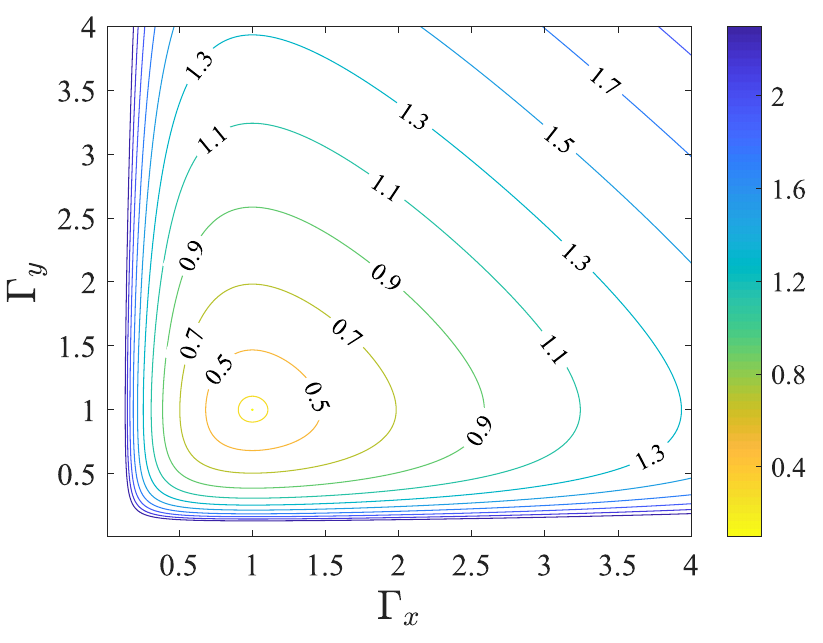} \caption{Contours of effective temperature $T_{\rm{eff}}$ with respect to $\Gamma_x$ and $\Gamma_y$ for $\omega_x = \omega_y = 0$. Here, we have set $\kappa_f$ = $\omega_z$ = $k_B$ = 1 for simplicity. $T_{\rm eff}$ takes the minimum, $T_{\rm{eff}} = 0$, at $\Gamma_x = \Gamma_y = 1$.}	
	\label{Teff}
\end{figure}

Equations~(\ref{eq:xs})\,--\,(\ref{eq:eta}) provide a guideline to realize the steady state of the system by the static external field, the simultaneous and continuous measurement, and the feedback control. One of the most important cases is the one in which the direction of the external magnetic field is along the $z$ axis, where the Hamiltonian of the system $\hat{H}_s$ is diagonal with $\omega_x = \omega_y = 0$ and $\omega_z \neq 0$. Then the steady state given by Eqs.~(\ref{eq:xs})\,--\,(\ref{eq:zs}) becomes  
\begin{align}
x_s &= 0,
\\
y_s &= 0,
\\
z_s &= \frac{k_2 - k_1}{k_1 + k_2 + 2k_3},
\label{eq:ss}
\end{align}
which is diagonal in the energy basis so that it can be identified as a thermal state with some effective temperature. This means that the simultaneous measurement and feedback control introduced above effectively serve as a heat bath \cite{A. Levy}. When we perform both the measurement and the feedback control on the system for a sufficiently long time, the system eventually reaches a thermal state. Since $z_s$ is independent of the sign of $\omega_z$ as seen in Eqs.~(\ref{eq:zs}) and~(\ref{eq:eta}), we can assume $\omega_z > 0$ without loss of generality, and the effective temperature $T_{\rm{eff}}$ can be obtained as 
\begin{align}
T_{\rm{eff}} = \frac{2 \omega_z}{k_B} \left( \ln \frac{k_1 + k_3}{k_2 + k_3}\right) ^{-1},
\label{eq:Teff}
\end{align}
where $k_B$ is the Boltzmann constant. Since $(k_1 + k_3)/(k_2 + k_3)$ is always larger than 1, the effective temperature $T_{\rm{eff}}$ is a monotonically increasing function of $\omega_z$, and $T_{\rm{eff}}$ satisfies $0 \leqslant T_{\rm{eff}} < \infty$. Thus, we can prepare the system in a diagonal steady state with an arbitrary positive effective temperature by setting proper $k_1$, $k_2$, $k_3$, and $\omega_z$ under $\omega_x = \omega_y = 0$. In order to see the dependence of $T_{\rm{eff}}$ on parameters $\Gamma_x$, $\Gamma_y$, and $\kappa_f$, we substitute Eqs.~(\ref{eq:k1})\,--\,(\ref{eq:k3}) into Eq.~(\ref{eq:ss}) and obtain 
\begin{align}
z_s = - 4 \left(  \frac{\Gamma_x}{\kappa_f} + \frac{\kappa_f}{\Gamma_x} +   
  \frac{\Gamma_y}{\kappa_f} + \frac{\kappa_f}{\Gamma_y} \right)  ^{-1}.
\end{align}
Here, $z_s$ first decreases when $\Gamma_x / \kappa_f$ and $\Gamma_y/ \kappa_f$ increase from $0$ to $1$, and then increases when $\Gamma_x / \kappa_f$ and $\Gamma_y/ \kappa_f$ increase from $1$ to infinity. $z_s$ takes the minimum at $\Gamma_x / \kappa_f = \Gamma_y/ \kappa_f =1$. According to Eq.~(\ref{eq:rho_s}), the probabilities in the excited state and the ground state are proportional to $(1 + z_s)$ and $(1 - z_s)$, respectively.  When $z_s$ decreases, the probability of the system in the excited state becomes smaller, which means that the effective temperature $T_{\rm{eff}}$ is lower. Therefore, $T_{\rm{eff}}$ first decreases and then increases when $\Gamma_x / \kappa_f$ and $\Gamma_y/ \kappa_f$ increase from $0$ to infinity, and $T_{\rm{eff}}$ takes the minimum value of zero at $\Gamma_x / \kappa_f = \Gamma_y/ \kappa_f = 1$. Figure~\ref{Teff} shows the contours of $T_{\rm{eff}}$ with respect to $\Gamma_x$ and $\Gamma_y$ for $\omega_x = \omega_y = 0$. 

Besides the final effective temperature, the timescale for relaxation to the asymptotic steady state is another important quantity. We first consider $z(t)$, which can be easily obtained from Eq.~(\ref{eq:zdot}) with $\omega_x = \omega_y = 0$:
\begin{align}
z(t) = \frac{k_2 -k_1}{k_1 + k_2 + 2k_3} + C_1 e^{-(k_1 + k_2 + 2k_3)t},
\label{eq:zt}
\end{align}
where $C_1$ is a constant determined by the initial condition $z(t = 0)$. From Eq.~(\ref{eq:zt}), the relaxation time $\tau_z$ of the spin-$z$ component can be defined as
\begin{align}
\tau_z \equiv & (k_1 + k_2 + 2k_3) ^{-1}
\notag
\\
= & 2 \left( \Gamma_x + \Gamma_y + \frac{\kappa_f ^2}{\Gamma_x} + \frac{\kappa_f ^2}{\Gamma_y}\right) ^{-1}.
\label{eq:tauz}
\end{align}
When $\Gamma_i \gg 1 $ or $\Gamma_i \ll \kappa_f ^2$ (with $i = x$ or $y$), the relaxation time $\tau_z$ is very short, which means the system can reach the steady state very quickly.

Next, we discuss $x(t)$ and $y(t)$. Equations~(\ref{eq:xdot}) and~(\ref{eq:ydot}) with $\omega_x=\omega_y=0$ can be written as
\begin{align}
\ddot{x} & = -(k_1 + k_2 + 2k_3) \dot{x} -\frac{\eta}{4(k_1 + k_2 + 2k_3)} x,
\label{eq:xddot}
\\
y & = -\frac{1}{2 \omega_z} \left( \dot{x} + \frac{k_1 + k_2}{2} x \right).
\label{eq:yt} 
\end{align}
The corresponding characteristic equation for Eq.~(\ref{eq:xddot}) is
\begin{align}
\mu ^2 - (k_1 + k_2 + 2k_3) \mu + \frac{\eta}{4(k_1 + k_2 + 2k_3)} = 0,
\label{eq:ce}
\end{align}
and the solution of Eq.~(\ref{eq:ce}) is:
\begin{align}
\mu_{\pm} = \frac{1}{2} \left[ (k_1 + k_2 + 2k_3)  \pm \sqrt{\Delta}\right]
\end{align}
with
\begin{align}
\Delta & \equiv (k_1 + k_2 + 2k_3)^2 - \frac{\eta}{k_1 + k_2 + 2k_3}
\notag
\\
& = 4k_3^2 - 16\omega_z^2.
\end{align}
The solutions of Eqs.~(\ref{eq:xddot}) and~(\ref{eq:yt}) are:
\begin{align}
x(t)  = & (C_2 + C_3 t )e^{-\frac{k_1 + k_2 + 2k_3}{2} t},
\\
y(t)  = & -\frac{1}{2 \omega_z}\left[  -k_3 C_2 + \left(  1   -k_3 t\right) C_3 \right]
e^{-\frac{k_1 + k_2 + 2k_3}{2} t} 
\end{align}
for $\Delta = 0$, and 
\begin{align}
x(t)  = & C_2 ^{\prime} e^{-\mu_{+} t} + C_3^{\prime} e^{-\mu_{-} t},
\\
y(t)  = & -\frac{1}{2 \omega_z} \left[  C_2^{\prime} \left(  \frac{k_1 + k_2}{2} -\mu_{+} \right)  e^{-\mu_{+} t} \right.
\notag
\\
& \left. +   C_3^{\prime} \left(  \frac{k_1 + k_2}{2} -\mu_{-} \right)   e^{-\mu_{-} t} \right] 
\end{align}
for $\Delta \neq 0$, where $C_2$, $C_3$, $C_2^{\prime}$, and $C_3^{\prime}$ are constants determined by the initial condition $x(t = 0)$ and $y(t = 0)$. When $\Delta \leqslant 0$, both $x(t)$ and $y(t)$ decay as $\exp{ \left(  -\frac{k_1 + k_2 + 2k_3}{2} t\right) }$. Thus, the relaxation time $\tau_x$ and $\tau_y$ of the $x$ and $y$ components of the spin can be defined as
\begin{align}
\tau_x = \tau_y \equiv & 2(k_1 + k_2 + 2k_3)^{-1}
\notag
\\
= & 2 \tau_z.
\end{align}
According to the previous discussion, the system reaches the steady state very quickly if  $\Gamma_i \gg 1 $ or $\Gamma_i \ll \kappa_f ^2$. When $\Delta > 0$, the decay rates of $x(t)$ and $y(t)$ are mainly determined by the term with $\exp{(-\mu_{-} t)}$ since $\mu_{-} < \mu_{+}$. Consequently, the definition of the relaxation time $\tau_x$ and $\tau_y$ is 
\begin{align}
\tau_x = \tau_y \equiv & \mu_{-}^{-1}
\notag
\\
= &  2 \left( k_1 + k_2 + 2k_3 - \sqrt{\Delta} \right) ^{-1}
\notag
\\
\leqslant &  2 \left( k_1 + k_2 + 2k_3 - \sqrt{4k_3^2} \right) ^{-1}
\notag
\\
= &  2 \left( k_1 + k_2 + 2k_3 - 2 \left| k_3 \right| \right) ^{-1}.\label{eq:tauxy}
\end{align}
When $k_3 \geqslant 0$, Eq.~(\ref{eq:tauxy}) reads
\begin{align}
\tau_x = \tau_y \leqslant &  2 (k_1 + k_2 )^{-1}
\notag
\\
= & 2 \left( \Gamma_y + \frac{\kappa_f^2}{\Gamma_x} \right) ^{-1}.
\end{align}
The relaxation time is very short if $\Gamma_y \gg 1$ or $\Gamma_x \ll \kappa_f^2$. When $k_3 < 0$ on the other hand, Eq.~(\ref{eq:tauxy}) reads
\begin{align}
\tau_x = \tau_y \leqslant &  2 (k_1 + k_2 + 4k_3)^{-1}
\notag
\\
= & 2 \left( \Gamma_x + \frac{\kappa_f^2}{\Gamma_y} \right) ^{-1},
\end{align}
and the relaxation time is very short if $\Gamma_x \gg 1$ or $\Gamma_y \ll \kappa_f^2$.
In summary, the relaxation time of the system is controllable by the strengths of the measurement and the feedback control, and the system can reach the steady state very quickly if $\Gamma_x, \Gamma_y \gg 1 $ or $\Gamma_x, \Gamma_y \ll \kappa_f ^2$.

\begin{figure}[tb!]
	\centering \includegraphics[scale=0.55]{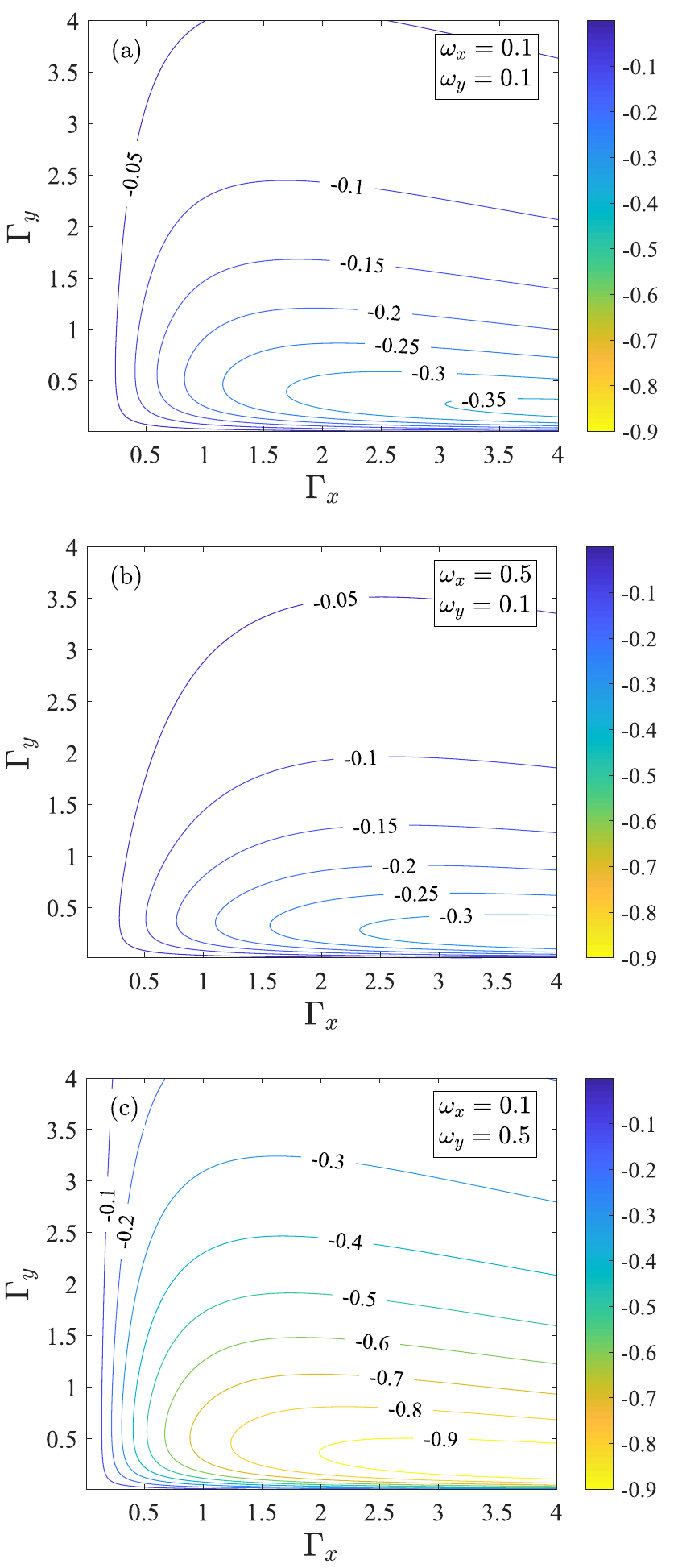}
	\caption{Contours of $x_s$ with respect to $\Gamma_x$ and $\Gamma_y$ at (a) $(\omega_x ,\omega_y ) = (0.1, 0.1)$, (b) $(\omega_x ,\omega_y ) = (0.5, 0.1)$, and (c) $(\omega_x ,\omega_y ) = (0.1, 0.5)$. Here we set $\omega_z = 0$ and $\kappa_f = 1$.} 	
	\label{sigx}
\end{figure}

Let us now discuss more general cases where the direction of the static external magnetic field is arbitrary, i.e., $\omega_x $, $\omega_y $, and $\omega_z $ are nonzero in general. In such cases, the Hamiltonian $\hat{H}_s$ contains off-diagonal terms and these off-diagonal terms generate coherence between the ground state and the excited state of the operator $\hat{\sigma}_z$, which leads to the situation in which the steady state is no longer a thermal state. To get a better understanding of the effect of the off-diagonal terms, we set $\omega_z = 0$ and keep $\omega_x$ and $\omega_y$ to be nonzero. Then $x_s$ and $y_s$, which characterize the effects of the off-diagonal terms on the steady state become 
\begin{align}
x_s &= 4\eta ^{-1} \omega_y (k_2 - k_1) (k_1 + k_2 + 4k_3),
\label{eq:x_s}
\\
y_s &= 4\eta ^{-1} \omega_x (k_1 - k_2) (k_1 + k_2 ).
\label{eq:y_s}
\end{align}

From Eqs.~(\ref{eq:x_s}) and~(\ref{eq:y_s}), we can clearly see that $x_s$ is more sensitive to $\omega_y$ than $\omega_x$ while $y_s$ is more sensitive to $\omega_x$ than $\omega_y$. Let us focus on $x_s$ as an example. Figure~\ref{sigx} shows $x_s$ as a function  of $\Gamma_x$ and $\Gamma_y$ for different values of $(\omega_x, \omega_y)$: $(0.1, 0.1)$ [Fig.~\ref{sigx}(a)],  $(0.5, 0.1)$ [Fig.~\ref{sigx}(b)], and $(0.1, 0.5)$ [Fig.~\ref{sigx}(c)]. One can clearly see that by comparing Figs.~\ref{sigx}(a) and~\ref{sigx}(b), the contours change only a little by changing $\omega_x$ with $\omega_y$ fixed, while by comparing Figs.~\ref{sigx}(a) and~\ref{sigx}(c), the contours drastically change by changing $\omega_y$ with $\omega_x$ fixed. Therefore, to control $x_s$ over a wide range, we should tune $\omega_y$ instead of $\omega_x$; on the other hand, to control $x_s$ accurately, we should first choose an appropriate value of $\omega_y$ and then tune $\omega_x$ with $\omega_y$ fixed.

\begin{figure}[tb!]
	\centering \includegraphics[scale=0.55]{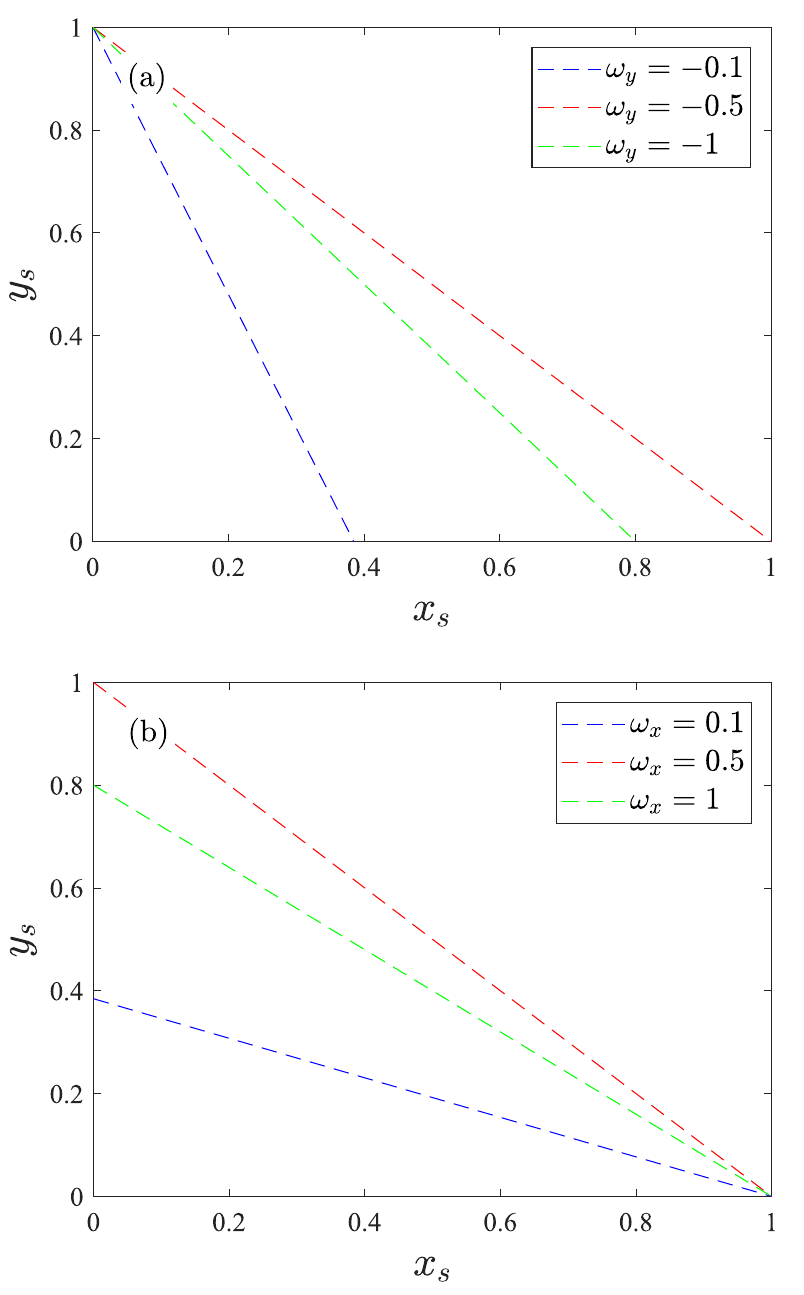}
	\caption{Boundaries of the region of possible $(x_s, y_s)$ for different values of $(\omega_x, \omega_y)$. Values of $(x_s, y_s)$ below the dashed line are realizable for each $(\omega_x, \omega_y)$. (a) $\omega_x = 0.5$ and $\omega_y = -0.1$ (blue dashed line), $-0.5$ (red dashed line), and $-1$ (green dashed line). (b) $\omega_y = - 0.5$ and $\omega_x = 0.1$ (blue dashed line), $0.5$ (red dashed line), and $1$ (green dashed line). Here, we set $\kappa_f = 1$, and negative values of $\omega_y$ to get positive $x_s$.}	
	\label{depend}
\end{figure}

Figure~\ref{sigx} also implies that the range of the accessible values of $x_s$ and $y_s$ by varying $\Gamma_x$ and $\Gamma_y$ depends on $\omega_x$ and $\omega_y$. To get a better understanding of this dependence, we show the region of possible values of $(x_s, y_s)$ for several different values of $(\omega_x, \omega_y)$ in Fig.~\ref{depend}. Figure \ref{depend}(a) shows the boundaries of this region for different values of $\omega_y=-0.1$, $-0.5$, and $-1$, with $\omega_x$ fixed at $0.5$. Note that these boundaries are straight lines within the numerical accuracy. With increasing the absolute value of $\omega_y$ from zero, the range of the possible values of $x_s$ first increases and becomes maximum at $|\omega_y|=0.5$, then decreases, while the range of the possible values of $y_s$ is unchanged from $0 \leqslant y_s \leqslant 1$. On the other hand, Fig.~\ref{depend}(b) shows the boundaries for different values of $\omega_x=0.1$, $0.5$, and $1$, with $\omega_y$ fixed at $-0.5$. The result is similar to that of Fig.~\ref{depend}(a), but $x_s$ and $y_s$ are switched: As $\omega_x$ increases from zero, the range of the possible values of $y_s$ first increases and becomes maximum at $\omega_x=0.5$, then decreases, while the range of the possible values of $x_s$ is unchanged from $0 \leqslant x_s \leqslant 1$. These results are consistent with the discussion in the last paragraph, and they also highlight the importance of choosing parameters $\omega_x$ and $\omega_y$ properly in preparing the state of the system. A given target steady state with some $x_s$ and $y_s$ is realizable only for parameters $(\omega_x, \omega_y)$ in some region.

\section{Conclusion \label{sec:conclusion}}

In summary, we have generalized the Arthurs-Kelly measurement model for a single-shot, simultaneous measurement to two arbitrary observables of the system whose commutator is not necessarily a $c$-number. We have found that this generalized measurement model is valid only when the coupling between the system and the detectors is sufficiently weak. By applying this generalized model to the continuous measurement of two arbitrary observables, we have derived both unconditioned and conditioned master equations. We have shown that the unconditioned master equation takes the Lindblad form in the continuous limit, even if the coupling is not infinitesimally small. Moreover, we have found that there is no effect of the interplay of the two measurements in the continuous limit, even if the two observables are noncommutative and their commutator is not a $c$-number. Finally, taking a spin-$1/2$ system as an example, we have illustrated that we can prepare a designated state as an asymptotic steady state of the time evolution by a static external field, the simultaneous, continuous measurement, and the feedback control based on the formalism derived in this work. We have obtained analytical expressions of the steady state and the timescale of the relaxation to the steady state, which offer a guiding principle for controlling the system. We have demonstrated that by appropriately setting the static external field and the strengths of the measurement and the feedback, we can control both the populations of the ground and the excited states and the coherence between them. Our results show that feedback control based on simultaneous, continuous measurement of multiple observables is one of the promising techniques which allows us to control the quantum state in a versatile manner.

\begin{acknowledgments}
We thank Luis Pedro García-Pintos and Peter Talkner for helpful discussions and comments. This work was supported by NSF of China (Grants No.~11975199 and No.~11674283), the Zhejiang Provincial Natural Science Foundation Key Project (Grant No.~LZ 19A050001), the Fundamental Research Funds for the Central Universities (Grants No.~2017QNA3005 and No.~2018QNA3004), and by the Zhejiang University 100 Plan.
\end{acknowledgments}


\begin{thebibliography}{99}
\bibitem{note:units} Depending on the choice of the quantities for $\hat{A}$ and $\hat{B}$, an appropriate set of additional universal physical constants are also set to be unity.
\bibitem{note:angularmom} Here, we consider a general angular momentum, not restricted to a spin-$1/2$, which is discussed in Sec.~\ref{sec:tls}.
\bibitem{note:covar} It is difficult to include the correlation between $\hat{x}_1$ and $\hat{x}_2$ within the present formalism. The crucial point is that the measurement outcomes $x_1(n)$ and $x_2(n)$, which are now described by two correlated Gaussian random variables, cannot be written with two independent It\^{o} increments $d\xi_1$ and $d\xi_2$ as Eqs.~(\ref{eq:out1}) and (\ref{eq:out2}).
\bibitem{note:explain} Here, $z_s$ is always negative for the choice of $\hat{F} = \hat{c}^{\dagger}$. If we choose $\hat{F} = -\hat{c}^{\dagger}$ instead, $z_s$ is always positive.
\end{thebibliography}
\end{document}